\documentclass[journal,comsoc]{IEEEtran}
\usepackage[T1]{fontenc}

\ifCLASSINFOpdf
  \usepackage[pdftex]{graphicx}
  \graphicspath{{../pdf/}{../jpeg/}}
  \DeclareGraphicsExtensions{.pdf,.jpeg,.png}
\else
  \usepackage[dvips]{graphicx}
  \graphicspath{{../eps/}}
  \DeclareGraphicsExtensions{.eps}
\fi

\usepackage{amsmath}
\usepackage{stfloats}
\usepackage{bm}
\usepackage{cite}

\newtheorem{theorem}{Theorem}

\newtheorem{proposition}[theorem]{Proposition}
\usepackage{algorithm}  
\usepackage{algorithmicx}  
\usepackage{algpseudocode}

\usepackage{url}

\begin{document}
\bstctlcite{BSTcontrol}

\title{Secure Wireless Transmission for Reconfigurable Intelligent Surface Aided Full Duplex Systems}
\author{
	\IEEEauthorblockN{Pengxin Guan, Yiru Wang and Yuping Zhao}
	\thanks{This work was supported by the National Key R\&D Program of China under Grant No.2020YFB1805102.	
	\emph{(Corresponding author: Yuping Zhao.)}}
\thanks{ Pengxin Guan, Yiru Wang and Yuping Zhao are with the School of
	Electronics Engineering and Computer Science, Peking University, Beijing
	100080, China. (email: guanpengxin@pku.edu.cn; yiruwang@stu.pku.edu.cn;
	yuping.zhao@pku.edu.cn).}
}

\markboth{Journal of \LaTeX\ Class Files,~Vol.~14, No.~8, August~2015}%
{Shell \MakeLowercase{\textit{et al.}}: Bare Demo of IEEEtran.cls for IEEE Communications Society Journals}

\maketitle

\begin{abstract}
This letter considers the secure communication in a reconfigurable intelligent surface (RIS) aided full duplex (FD) system. An FD base station (BS) serves an uplink (UL) user and a downlink (DL) user simultaneously over the same time-frequency dimension assisted by a RIS in the presence of an eavesdropper. In addition, the artificial noise (AN) is also applied to interfere the eavesdropper's channel. We aim to maximize the sum secrecy rate of UL and DL users by jointly optimizing the transmit beamforming, receive beamforming and AN covariance matrix at the BS, and passive beamforming at the RIS. To handle the non-convex problem, we decompose it into tractable subproblems and propose an efficient algorithm based on alternating optimization framework. Specifically, the receive beamforming is derived as a closed-form solution while other variables are obtained by using semidefinite relaxation (SDR) method and successive convex approximation (SCA) algorithm. Simulation results demonstrate the superior performance of our proposed scheme compared to other baseline schemes.
\end{abstract}


%
\IEEEpeerreviewmaketitle

\section{Introduction}

\IEEEPARstart{R}{econfigurable} intelligent surface (RIS) has been recognized as a key technology for 6G since it is capable of improving spectrum efficiency (SE) for wireless communication systems \cite{r1}. Recently, RIS is also implemented to enhance the physical layer security (PLS) performance \cite{r2,r3,r4,r5}. By intelligently configuring the phase shifts at the RIS, which is also called passive beamforming, the signal power can be enchanced at legitimate receivers and attenuated at eavesdropper. In \cite{r2} and \cite{r3}, the authors maximized secrecy rate by jointly optimizing the active and passive beamforming in a RIS-aided system. To further improve the PLS performance, artificial noise (AN) is also applied to the RIS assisted communication systems \cite{r4,r5}.

However, the literature mentioned above only considered the half-duplex (HD) communication systems, which cause the SE loss. The full-duplex (FD) technology enables signal transmission and reception over the same time-frequency dimension and thus can improve the SE compared with HD \cite{r6}. The RIS-aided FD systems have gained some attention, such as sum rate maximization \cite{r7} and transmit power minimization \cite{r8}. Nevertheless, the secure communication in RIS-aided FD systems has not been well studied. The authors in \cite{r9} studied the robust secure beamforming in a RIS-aided point-to-point FD system. The authors in \cite{r10} used the deep reinforcement learning to maximize the sum secrecy rate (SSR) in a RIS assisted FD system. However, the AN, which has been shown to enhance system security, was not considered in these works. In addition, the authors in \cite{r11} only optimized the uplink secrecy rate optimization in a RIS aided FD system, while both uplink and downlink security needs to be considered in FD systems generally \cite{r12,r13}.
\begin{figure}
	\centering
	\includegraphics[width=2.9in]{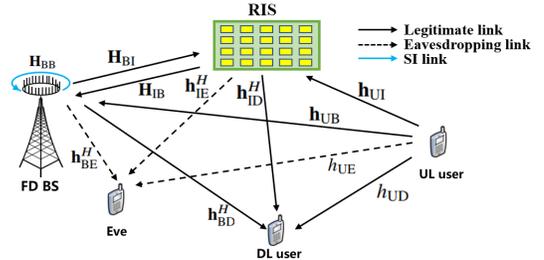}
	\caption{Illustration of the RIS aided FD secure system.}
\end{figure}

In this letter, we consider a RIS aided FD system, where a FD base station (BS) communicates with an uplink (UL) user and a downlink (DL) user simultaneously in the presence of an eavesdropper. We aim to maximize the SSR by jointly optimizing the transmit beamforming, receive beamforming and AN covariance matrix at the BS, and passive beamforming at the RIS. The main contributions are summarized as follows:
\begin{itemize}
	\item We formulate a SSR maximization problem, subjected to the maximum transmit power of the BS and the constant modulus constraints of RIS elements.
	\item We decompose the non-convex problem into tractable subproblems and propose an efficient algorithm based on alternating optimization (AO) framework.
	\item Simulation results show the benefits of our proposed scheme compared with other benchmark schemes.
\end{itemize}

\emph{Notation}: $\left| x \right|$ and arg($x$) are the absolute value of $x$ and phase of a complex number $x$. $\left\| {{{\bf{x}}}} \right\|$ is the Euclidean norm of ${\bf{x}}$. ${\left[ {\bf{x}} \right]_{m}}$ is the $m$-th element of ${\bf{x}}$. Tr(${\bf{X}}$), ${{\bf{X}}^H}$, rank(${\bf{X}}$) and  ${\left[ {\bf{X}} \right]_{i,i}}$ denote the trace, conjugate transpose, rank and ($i,i$)-entry of the matrix ${\bf{X}}$, respectively. diag(${\bf{x}}$) is a diagonal matrix with the entries of ${\bf{x}}$ on its main diagonal. ${\bf{X}} \succeq {\bf{0}}$ denotes that ${\bf{X}}$ is positive semidefinite. $\nabla$ represents the gradient operation.
\section{SYSTEM MODEL AND PROBLEM FORMULATION}

\subsection{System Model Description}
As shown in Fig. 1, we consider a RIS-aided FD secure communication system, which consists of an FD BS, an UL user, a DL user, a RIS and an eavesdropper (Eve). The FD BS is equipped with ${N_\text{T}}$ transmit antennas and ${N_\text{R}}$ receive antennas, while the two users and Eve are equipped with a single antenna. The RIS with $M$ passive elements is deployed to improve system security performance.

The transmitted signal from the BS is given by
\begin{equation}
	\setlength{\abovedisplayskip}{2pt}
\setlength{\belowdisplayskip}{2pt}
{\bf{x = }}{{\bf{w}}}{{{s}}_{\text{D}}}{\bf{ + v}},
\end{equation}
where  ${{\bf{w}}} \in {{\mathbb C}^{{N_\text{T}} \times 1}}$ is the transmit beamforming at the BS and $s_\text{D}$ is the DL information symbol with normalized power. In addition, ${{\bf{v}}} \in {{\mathbb C}^{{N_\text{T}} \times 1}}$$\sim {\mathcal{CN}} (0,{\bf{V}})$ is the AN vector and $ {\bf{V}} $ is the corresponding covariance matrix.

Denote ${{\bf{H}}_\text{BI}} \in {{\mathbb C}^{M \times {N_\text{T}}}}$, ${\bf{H}}_\text{IB} \in {{\mathbb C}^{{N_\text{R}} \times M}}$, ${{\bf{h}}_\text{ID}^{H}} \in {{\mathbb C}^{1 \times M}}$, ${\bf{h}}_\text{IE}^H \in {{\mathbb C}^{1 \times M}}$, ${{\bf{h}}_\text{BE}^{H}} \in {{\mathbb C}^{1 \times {N_\text{T}} }}$, ${{\bf{h}}_\text{BD}^{H}} \in {{\mathbb C}^{1 \times {N_\text{T}}}}$, ${{{h}}_\text{UE}} \in {{\mathbb C}}$, ${{{h}}_\text{UD}} \in {{\mathbb C}}$, ${{\bf{h}}_\text{UB}} \in {{\mathbb C}^{{N_\text{R}} \times 1}}$, ${{\bf{h}}_\text{UI}} \in {{\mathbb C}^{{M} \times 1}}$ as the channel from the BS to the RIS, from the RIS to the BS, from the RIS to the DL user, from the RIS to the Eve, from BS to the Eve, from the BS to the DL user, from the UL user to the Eve, from the UL user to the DL user, from the UL user to the BS, and from the UL user to the RIS, respectively. ${{\bf{H}}_\text{BB}\in {{\mathbb C}^{{N_\text{R}} \times {N_\text{T}}}}}$ represents the residual SI channel at the BS \cite{r11}. ${\bf{\Theta}}  = {\rm{diag}}\left\{ {\left[ {{\phi _1}, {\phi _2}, \cdots ,{\phi _M}} \right]} \right\}$ is the phase-shift matrix of the RIS and ${\phi _m} = {e^{j{\theta _m}}}$, where ${\theta _m}\in\left[ {0,2\pi } \right)$ denotes the phase-shift of the $m$-th reflecting element. Similar to \cite{r2,r3,r4}, we assume full channel state information of the system is available.

Then, the signal received at the BS can be expressed as
\begin{equation}
	\setlength{\abovedisplayskip}{2pt}
\setlength{\belowdisplayskip}{2pt}
{{y}_\text{U}}{\bf{=w}}_{\text{R}}^{{H}}{\bf{(}}{{\bf{H}}_{{\text{IB}}}}{\bf{\Theta}}{{\bf{h}}_{{\text{UI}}}}{\bf{ + }}{\bf{h}_{\text{UB}}}{\bf{)}}\sqrt {{{{P}}_{\text{U}}}} {{{s}}_{\text{U}}}{\bf{ + w}}_{\text{R}}^{{H}}{\bf{}}{{\bf{H}}_{{\text{BB}}}}{\bf{x + }}{\bf{ w}}_{\text{R}}^{{H}}{\bf{}}{{\bf{n}}_{\text{B}}}{\bf{}},
\end{equation}
where $s_\text{U}$ and ${P_{\text{U}}}$ are the corresponding transmitted data symbol with normalized power and transmit power of the UL user, respectively. ${\bf{w}}_{\rm{R}}^{{H}}$ is the receive beamforming and ${{\bf{n}}_{\rm{B}}} \sim \mathcal{CN}(0,\sigma _{\rm{B}}^2{{\bf{I}}_{{N_\text{R}}}})$ is the additive white Gaussian noise (AWGN) at the BS. We neglect the reflected SI signal by RIS because it is much weaker than the SI at the BS \cite{r7}.

The signal received by the DL user and Eve are given by
\begin{equation}
	\setlength{\abovedisplayskip}{2pt}
\setlength{\belowdisplayskip}{2pt}
{y_\text{D}} = ({\bf{h}}_{{\text{ID}}}^{{H}}{\bf{\Theta}}{{\bf{H}}_\text{BI}} + {\bf{h}}_{{\text{BD}}}^{{H}}){\bf{x}} + ({\bf{h}}_{{\text{ID}}}^{{H}}{\bf{\Theta}}{{\bf{h}}_{{\text{UI}}}}{\bf{ + }}{{{h}}_{{\text{UD}}}})\sqrt {{{{P}}_{\text{U}}}} {{{s}}_{\text{U}}} + {n_\text{D}},
\end{equation}
and
\begin{equation}
	\setlength{\abovedisplayskip}{2pt}
\setlength{\belowdisplayskip}{2pt}
{y_\text{E}} = ({\bf{h}}_{{\text{IE}}}^{{H}}{\bf{\Theta}}{{\bf{h}}_{{\text{UI}}}} + {{{h}}_{{\text{UE}}}})\sqrt {{{{P}}_{\text{U}}}} {{{s}}_{\text{U}}} + ({\bf{h}}_{{\text{IE}}}^{{H}}{\bf{\Theta}}{{\bf{H}}_\text{BI}} + {\bf{h}}_{{\text{BE}}}^{{H}}){\bf{x}} + {n_\text{E}},
\end{equation}
where ${n_{\rm{D}}}\sim \mathcal{CN}(0,\sigma _{\rm{D}}^2)$ and ${n_{\rm{E}}} \sim \mathcal{CN}(0,\sigma _{\rm{E}}^2)$ are the AWGN at the DL user and Eve, respectively.

Further, the signal-to-interference-plus-noise ratio (SINR) at the UL and DL users can be derived as
\begin{equation}
	\setlength{\abovedisplayskip}{2pt}
\setlength{\belowdisplayskip}{2pt}
{\gamma _{\text{U}}} = \frac{{{P_{\text{U}}}{{\left| {{\bf{w}}_{\text{R}}^H({{\bf{H}}_{{\text{IB}}}}{\bf{\Theta}}{{\bf{h}}_{{\text{UI}}}}{\bf{ + }}{{\bf{h}}_{{\text{UB}}}})} \right|}^2}}}{{{{\left| {{\bf{w}}_{\text{R}}^H{{\bf{H}}_{{\text{BB}}}}{\bf{w}}} \right|}^2} + {\bf{w}}_{\rm{R}}^H{{\bf{H}}_{{\text{BB}}}}{\bf{VH}}_{{\text{BB}}}^H{{\bf{w}}_{\text{R}}} + \sigma _{\text{B}}^2{{\left\| {{\bf{w}}_{\text{R}}} \right\|}^2}}}
\end{equation}
and
\begin{equation}
	\setlength{\abovedisplayskip}{2pt}
\setlength{\belowdisplayskip}{2pt}
{\gamma _{\rm{D}}} = \frac{{{{\left| {{\bf{\hat h}}_{{\rm{BD}}}^H{\bf{w}}} \right|}^2}}}{{{P_{\rm{U}}}{{\left| {{\bf{h}}_{{\rm{ID}}}^H{\bf{\Theta}}{{\bf{h}}_{{\rm{UI}}}}{\bf{ + }}{h_{{\rm{UD}}}}} \right|}^2} + {\bf{\hat h}}_{{\rm{BD}}}^H{\bf{V}}{{{\bf{\hat h}}}_{{\rm{BD}}}} + \sigma _{\rm{D}}^2}},
\end{equation}
where ${\bf{\hat h}}_{{\rm{BD}}}^H = {\bf{h}}_{{\rm{ID}}}^H{\bf{\Theta}}{{\bf{H}}_{{\rm{BI}}}}{\bf{ + h}}_{{\rm{BD}}}^H$ for simplification.

We assume the Eve decodes the DL and UL symbol separately, as in \cite{r12,r13}. Then, the SINRs of the DL and UL signals at the Eve can be expressed as
\begin{equation}
	\setlength{\abovedisplayskip}{2pt}
\setlength{\belowdisplayskip}{2pt}
\gamma _{\rm{D}}^{\rm{E}} = \frac{{{{\left| {{\bf{\hat h}}_{{\rm{BE}}}^H{\bf{w}}} \right|}^2}}}{{{P_{\rm{U}}}{{\left| {{\bf{h}}_{{\rm{IE}}}^H{\bf{\Theta}}{{\bf{h}}_{{\rm{UI}}}}{\bf{ + }}{h_{{\rm{UE}}}}} \right|}^2} + {\bf{\hat h}}_{{\rm{BE}}}^H{\bf{V}}{{{\bf{\hat h}}}_{{\rm{BE}}}} + \sigma _{\rm{E}}^{\rm{2}}}},
\end{equation}
and
\begin{equation}
	\setlength{\abovedisplayskip}{4pt}
\setlength{\belowdisplayskip}{4pt}
\gamma _{\rm{U}}^{\rm{E}} = \frac{{{P_{\rm{U}}}{{\left| {{\bf{h}}_{{\rm{IE}}}^H{\bf{\Theta}}{{\bf{h}}_{{\rm{UI}}}}{\bf{ + }}{h_{{\rm{UE}}}}} \right|}^2}}}{{{{\left| {{\bf{\hat h}}_{{\rm{BE}}}^H{\bf{w}}} \right|}^2} + {\bf{\hat h}}_{{\rm{BE}}}^H{\bf{V}}{{{\bf{\hat h}}}_{{\rm{BE}}}} + \sigma _{\rm{E}}^{\rm{2}}}},
\end{equation}
where ${\bf{\hat h}}_{{\rm{BE}}}^H = {\bf{h}}_{{\rm{IE}}}^H{\bf{\Theta}}{{\bf{H}}_{{\rm{BI}}}}{\bf{ + h}}_{{\rm{BE}}}^H$.

According to \cite{r12} and \cite{r13}, the security rate of both links are defined as ${R_i^{\text{Sec}}} = {\left[ {{{\log }_2}(1 + {\gamma _i}) - {{\log }_2}(1 + \gamma _i^\text{E})} \right]^ + },i = \{ \text{U}, \text{D}\}$, where ${\left[  \cdot  \right]^ + } = \max \{  \cdot ,0\} $.
\subsection{Problem Formulation}
In this work, we aim to maximize the SSR by jointly optimizing the transmit beamforming, receive beamforming, AN covariance matrix and passive beamforming. Mathematically, the optimization problem is formulated as
\begin{subequations}\label{opt1}
	\setlength{\abovedisplayskip}{2pt}
	\setlength{\belowdisplayskip}{2pt}
	\begin{alignat}{2} 
		\mathcal{P}1:& \mathop {\max }\limits_{\bf{w},\bf{w_R},\bf{\Theta}, \bf{V}} &\ &\sum\limits_{i = \{ \text{U, D}\} }  {{R_i^{\text{Sec}}}}  \\
		&\;\quad{\textrm {s.t.}}&&\text{Tr}({{\bf{w}}}{\bf{w}}^{{H}} + {\bf{V}}) \le {P_{\max }},\label{opt1B}\\
		&&&{\left\| {{{\bf{w}}_{\text{R}}}} \right\|^2} = 1,\\
		&&&{{\bf{V}}} \,\succeq {\bf{0}},\\
		&&&\left| {{\phi _m}} \right| = 1,\;{\rm{ }}m = 1, \cdots ,M,
	\end{alignat}
\end{subequations}	
where ${P_{\max }}$ is the maximum transmit power at the BS and (9e) imposes a unit modulus constraint on the RIS.

Note that the operator ${\left[  \cdot  \right]^ + }$ has no impact on the optimization and hence it will be omitted in subsequent analysis\footnote{If the achievable secrecy rate on a certain communication link is negative, we can turn off the transmission of this link and apply our proposed algorithm to complete the system optimization again, as in \cite{r14}.}. Problem $\mathcal{P}1$ is still challenging to solve due to the non-convex objective function and unit modulus constraint, and then we will propose an efficient AO algorithm in subsequent section.
\section{SUM SECRECY RATE MAXIMIZATION ALGORITHM DESIGN}
\subsection{Optimizing the Receive Beamforming Vector ${{\bf{w}}_{\rm{R}}}$}
In this subsection, we aim to optimize the receive beamforming ${{\bf{w}}_{\rm{R}}}$ with fixed ${{\bf{w}}}$, ${{\bf{v}}}$ and ${{\bf{\Theta}}}$. Note that ${{\bf{w}}_{\rm{R}}}$ is only related to the UL SINR at the BS, thus the receive beamforming optimization problem can be expressed as
\begin{subequations}\label{opt1}
		\setlength{\abovedisplayskip}{2pt}
	\setlength{\belowdisplayskip}{2pt}
	\begin{alignat}{2} 
	\mathcal{P}2:& \;\,\mathop {\max }\limits_{\bf{w_R}} &\ &{\gamma _\text{U}}  \label{opt1A} \\
	& \quad{\textrm {s.t.}}&&{\left\| {{{\bf{w}}_{\text{R}}}} \right\|^2} = 1.
	\end{alignat}
\end{subequations}	

The problem $\mathcal{P}2$ is a generalized eigenvalue problem. Let ${{\bf{W}}}{\bf{ = }}{{\bf{w}}}{\bf{w}}^{{H}}$ and then the optimal $\bf{w_R}$ can be given as \cite{r11}
\begin{equation}
	\setlength{\abovedisplayskip}{4pt}
\setlength{\belowdisplayskip}{4pt}
{\bf{w}}_\text{R}^* = \frac{{{{\left[ {{{\bf{H}}_{{\text{BB}}}}{\bf{(}}{{\bf{W}}}{\bf{ + V)H}}_{{\text{BB}}}^{{H}}{\bf{ + \sigma }}_{\text{B}}^{\bf{2}}{\bf{I}}_{{N}_\text{R}}} \right]}^{{\bf{ - 1}}}}{\bf{(}}{{\bf{H}}_{{\text{IB}}}}{\bf{\Theta}}{{\bf{h}}_{{\text{UI}}}}{\bf{ + }}{{\bf{h}}_{{\text{UB}}}}{\bf{)}}}}{{\left\| {{{\left[ {{{\bf{H}}_{{\text{BB}}}}{\bf{(}}{{\bf{W}}}{\bf{ + V)H}}_{{\text{BB}}}^{{H}}{\bf{ + \sigma }}_{\text{B}}^{\bf{2}}{\bf{I}}_{{N}_\text{R}}} \right]}^{{\bf{ - 1}}}}{\bf{(}}{{\bf{H}}_{{\text{IB}}}}{\bf{\Theta}}{{\bf{h}}_{{\text{UI}}}}{\bf{ + }}{{\bf{h}}_{{\text{UB}}}}{\bf{)}}} \right\|}}.
\end{equation}
\subsection{Optimizing the Transmit Beamforming Vector $\bf{w}$ and AN Covariance Matrix $\bf{V}$ }
In this subsection, we focus on the optimization of $\bf{w}$ and $\bf{V}$, while fixing the other variables. To facilitate formulation, we define ${{\bf{h}}_1} = \left( \begin{array}{l}
{\rm{diag}}({\bf{w}}_{\rm{R}}^H{{\bf{H}}_{{\rm{IB}}}}){{\bf{h}}_{{\rm{UI}}}}\\
{\bf{w}}_{\rm{R}}^H{{\bf{h}}_{{\rm{UB}}}}
\end{array} \right)$, ${{\bf{h}}_2} = {\bf{H}}_{{\rm{BB}}}^H{{\bf{w}}_{\rm{R}}}$, ${{\bf{h}}_3} = \left( \begin{array}{l}
{\rm{diag}}({\bf{h}}_{{\rm{ID}}}^H){{\bf{H}}_{{\rm{BI}}}}\\
{\bf{h}}_{{\rm{BD}}}^H
\end{array} \right)$, ${{\bf{h}}_4} = \left( \begin{array}{l}
{\rm{diag}}({\bf{h}}_{{\rm{ID}}}^H){{\bf{h}}_{{\rm{UI}}}}\\
{h_{{\rm{UD}}}}
\end{array} \right)$, ${{\bf{h}}_5} = \left( \begin{array}{l}
{\rm{diag}}({\bf{h}}_{{\rm{IE}}}^H){{\bf{H}}_{{\rm{BI}}}}\\
{\bf{h}}_{{\rm{BE}}}^H
\end{array} \right)$, ${{\bf{h}}_6} = \left( \begin{array}{l}
{\rm{diag}}({\bf{h}}_{{\rm{IE}}}^H){{\bf{h}}_{{\rm{UI}}}}\\
{h_{{\rm{UE}}}}
\end{array} \right)$, ${{\bf{H}}_1} = {{\bf{h}}_1}{\bf{h}}_1^H$, ${{\bf{H}}_2} = {{\bf{h}}_2}{\bf{h}}_2^H$, ${{\bf{H}}_4} = {{\bf{h}}_4}{\bf{h}}_4^H$, ${{\bf{H}}_6} = {{\bf{h}}_6}{\bf{h}}_6^H$ and ${\bf{q}} = {[{\phi _1}, \cdots ,{\phi _M},1]^H}$ and ${\bf{Q}} = {\bf{q}}{{\bf{q}}^H}$. Then, after some intuitive operations, we can formulate transmit beamforming vector and AN covariance matrix optimization problem as 
\begin{figure*}[hb]\hrulefill
		\setlength{\abovedisplayskip}{2pt}
	\setlength{\belowdisplayskip}{2pt}
		\begin{equation}
		\begin{array}{l}
		F({\bf{W}},{\bf{V}},{\bf{Q}}) = {\log _2}\{ {P_{\rm{U}}}{\rm{Tr}}({\bf{Q}}{{\bf{H}}_1}) + {\rm{Tr[}}{{\bf{H}}_2}({\bf{W}} + {\bf{V}})] + \sigma _{\rm{B}}^2\}  + {\log _2}\{ {P_{\rm{U}}}{\rm{Tr}}({\bf{Q}}{{\bf{H}}_4}) + {\rm{Tr[}}{\bf{h}}_3^H{\bf{Q}}{{\bf{h}}_3}({\bf{W}} + {\bf{V}})] + \sigma _{\rm{D}}^2\}  \\
		\quad\quad\quad\quad\quad\quad+{\log _2}[{P_{\rm{U}}}{\rm{Tr}}({\bf{Q}}{{\bf{H}}_6}) + {\rm{Tr(}}{\bf{h}}_5^H{\bf{Q}}{{\bf{h}}_5}{\bf{V}}) + \sigma _{\rm{E}}^2] + {\log _2}\{ {\rm{Tr[}}{\bf{h}}_5^H{\bf{Q}}{{\bf{h}}_5}({\bf{W}} + {\bf{V}})] + \sigma _{\rm{E}}^2\} .
		\end{array}
		\end{equation}
\end{figure*}
\begin{figure*}[hb]
		\setlength{\abovedisplayskip}{2pt}
	\setlength{\belowdisplayskip}{2pt}
	\begin{equation}
	\begin{array}{l}
	G({\bf{W}},{\bf{V}},{\bf{Q}})= {\log _2}\{ {\rm{Tr[}}{{\bf{H}}_2}({\bf{W}} + {\bf{V}})] + \sigma _{\rm{B}}^2\}  + {\log _2}[{P_{\rm{U}}}{\rm{Tr}}({\bf{Q}}{{\bf{H}}_4}) + {\rm{Tr(}}{\bf{h}}_3^H{\bf{Q}}{{\bf{h}}_3}{\bf{V}}) + \sigma _{\rm{D}}^2]\\
	\quad \quad\quad\quad\quad\quad+ 2{\log _2}\{ {P_{\rm{U}}}{\rm{Tr}}({\bf{Q}}{{\bf{H}}_6}) + {\rm{Tr[}}{\bf{h}}_5^H{\bf{Q}}{{\bf{h}}_5}({\bf{W}} + {\bf{V}})] + \sigma _{\rm{E}}^2\}.
	\end{array}
	\end{equation}
\end{figure*}
\begin{subequations}\label{opt1}
		\setlength{\abovedisplayskip}{4pt}
	\setlength{\belowdisplayskip}{4pt}
	\begin{alignat}{2} 
	\mathcal{P}3:& \;\,\mathop {\max }\limits_{\bf{W},\bf{V}} &\ &{F({{\bf{W}}}{\bf{,V}}) - G({{\bf{W}}}{\bf{,V}})}  \label{opt1A} \\
	& \quad{\textrm {s.t.}}&&\rm{rank}({{\bf{W}}}) = 1,\\
	&&&\text{Tr}({{\bf{W}}}+ {\bf{V}}) \le {P_{\max }},\\
	&&&{{\bf{W}}} \succeq {\bf{0}},\\
	&&&\,{{\bf{V}}} \succeq {\bf{0}},
	\end{alignat}
\end{subequations}	
where $F({{\bf{W}}}{\bf{,V}})$ and $G({{\bf{W}}}{\bf{,V}})$ are the simplified notations of $F({{\bf{W}}}{\bf{,V}}{\bf{,Q}})$ and $G({{\bf{W}}}{\bf{,V}}{\bf{,Q}})$ with given $\bf{Q}$, which are shown at the bottom of this page. Note that $F({{\bf{W}}}{\bf{,V}})$ and $G({{\bf{W}}}{\bf{,V}})$ are both concave with respect to $\bf{W}$ and $\bf{V}$, and thus (14a) is a difference-of-concave (DC) function. The problem $\mathcal{P}3$ is difficult to solve due to the non-convex objective function and rank-one constraint. However, we can utilize the first-order Taylor expansions to approximate $G({{\bf{W}}}{\bf{,V}})$ around local point (${\bf{W}}^{(n)}$,${\bf{V}}^{(n)}$) at the $n$-th iteration, which can be written as
\begin{equation}
	\setlength{\abovedisplayskip}{4pt}
\setlength{\belowdisplayskip}{4pt}
\begin{array}{l}
G({{\bf{W}}}{\bf{,V}}) \le G({\bf{W}}^{(n)}{\bf{,}}{{\bf{V}}^{(n)}}) + \text{Tr}[\nabla _{_{{\bf{W}}^{(n)}}}^HG({{\bf{W}}}{\bf{,V}})({{\bf{W}}} - {\bf{W}}^{(n)})]\\ 
\quad\quad\quad+ \text{Tr}[\nabla _{{{\bf{V}}^{(n)}}}^HG({{\bf{W}}}{\bf{,V}})({\bf{V - }}{{\bf{V}}^{(n)}})]= G({{\bf{W}}}{\bf{,V}}|{\bf{W}}^{(n)}{\bf{,}}{{\bf{V}}^{(n)}}),
\end{array}
\end{equation}
where $G({{\bf{W}}}{\bf{,V}}|{\bf{W}}^{(n)}{\bf{,}}{{\bf{V}}^{(n)}})$ is a global upper bound of $G({{\bf{W}}}{\bf{,V}})$. By dropping the rank-one constraint, the approximated problem at the $n$-th iteration can be reformulated as
\begin{subequations}\label{opt7}
		\setlength{\abovedisplayskip}{2pt}
	\setlength{\belowdisplayskip}{2pt}
	\begin{alignat}{2} 
	\mathcal{P}3^{\prime}:&\;\, \mathop {\max }\limits_{\bf{W},\bf{V}} &\ & F({{\bf{W}}}{\bf{,V}}) - G({{\bf{W}}}{\bf{,V}}|{\bf{W}}^{(n)}{\bf{,}}{{\bf{V}}^{(n)}})  \\
	&\quad{\textrm {s.t.}}&&\text{(14c)-(14e)}.\label{opt5B}
	\end{alignat}
\end{subequations}	

Problem $\mathcal{P}3^{\prime}$ is a convex optimization problem, and thus CVX can be used to obtain the optimal solution ${\bf{W}}^{(n+1)}$ and ${\bf{V}}^{(n+1)}$. By iteratively solving problem $\mathcal{P}3^{\prime}$ based on the successive convex approximation (SCA) method, the objective function in (14a) will increase and converges \cite{r15}. 

\begin{proposition}
	The optimal transmit beamforming matrix of the Problem $\mathcal{P}3^{\prime}$ satisfies $\rm{rank}({{\bf{W}^{*}}}) \le 1$.
\end{proposition}
\begin{IEEEproof}
	Please refer to the Appendix.
\end{IEEEproof}

According to the Proposition 1, when the BS transmit the DL information, i.e., $\bf{W}\ne\bf{0}$, we get $\text{rank}({{\bf{W}}^*})=1$ and then the eigenvalue decomposition (EVD) can be applied to recover the transmit beamforming vector $\bf{w}^{*}$.
\subsection{Optimizing the Phase Shift Matrix $\bf{\Theta}$}
Now, we investigate the optimization of phase shift matrix $\bf{\Theta}$ with the other variables given. The phase-shifts of RIS optimization problem can be rewritten as
\begin{subequations}\label{opt7}
		\setlength{\abovedisplayskip}{4pt}
	\setlength{\belowdisplayskip}{4pt}
	\begin{alignat}{2} 
	\mathcal{P}4:& \;\, \mathop {\max }\limits_{\bf{Q}}& &\ F({\bf{Q}}) - G({\bf{Q}})  \\
	&\quad{\textrm {s.t.}}&&\;{{\left[\bf{Q}\right]}_{m,m}} = 1,\;{\rm{ }}m = 1, \cdots ,M + 1 ,\label{opt6D}\\
	&&&\;{{\bf{Q}}} \succeq {\bf{0}},\\
	&&&\;\rm{rank}({{\bf{Q}}}) = 1,
	\end{alignat}
\end{subequations}	
where $F({\bf{Q}})$ and $ G({\bf{Q}})$ are the simplified notations of $F({{\bf{W}}}{\bf{,V}}{\bf{,Q}})$ and $G({{\bf{W}}}{\bf{,V}}{\bf{,Q}})$ with given $\bf{W}$ and $\bf{V}$. Note that $F({\bf{Q}})$ and $ G({\bf{Q}})$ are both concave with respect to $\bf{Q}$. Then, similiar to the optimization of transmit beamforming and AN covariance matrix, we can apply the SCA algorithm to solve this DC problem. Specifically, by using the first-order Taylor approximation for the concave function $ G({\bf{Q}})$, we have
\begin{equation}
	\setlength{\abovedisplayskip}{3pt}
\setlength{\belowdisplayskip}{3pt}
\begin{array}{l}
G({\bf{Q}}) \le G({{\bf{Q}}^{(i)}}) + \text{Tr}[\nabla _{_{{{\bf{Q}}^{(i)}}}}^HG({\bf{Q}})({\bf{Q}} - {{\bf{Q}}^{(i)}})]\\
 \quad\quad\;\;=G({\bf{Q}}|{{\bf{Q}}^{(i)}}),
\end{array}
\end{equation}
where ${{\bf{Q}}^{(i)}}$ is the solution at the $i$-th iteration. By dropping the rank-one constraint based on SDR method \cite{r2}, we get
\begin{subequations}\label{opt8}
		\setlength{\abovedisplayskip}{3pt}
	\setlength{\belowdisplayskip}{3pt}
	\begin{alignat}{2} 
	\mathcal{P}4^{\prime}:&\; \mathop {\max }\limits_{\bf{Q}} & &\ F({\bf{Q}}) - G({\bf{Q}}|{{\bf{Q}}^{(i)}})  \\
	&\quad{\textrm {s.t.}}&&\;\text{(17b)-(17c)}.
	\end{alignat}
\end{subequations}

The problem $\mathcal{P}4^{\prime}$ is a convex semidefinite program which can be solved directly by exploiting the convex optimization tool, e.g., CVX. Using the SCA algorithm to iteratively solve the problem $\mathcal{P}4^{\prime}$, the objective function in (17a) will increase and converges \cite{r15}. Then, we can apply the Gaussian randomization method to get a high-quality solution $\bf{q}^{*}$ \cite{r2}. Further, the phase-shifts of the RIS can be given by
\begin{equation}
	\setlength{\abovedisplayskip}{2pt}
\setlength{\belowdisplayskip}{2pt}
{{\bm{\theta}}^{*}_{m}} = -{\text{arg}({\bf{q}}_m^*/{\bf{q}}_{M + 1}^*)},{\rm{ }}\;m = 1, \cdots ,M.
\end{equation}

The corresponding phase shift matrix can be expressed as ${\bf{\Theta}}^{*}=\text{diag}\{e^{j\bm{\theta}^{*}}\}$. However, the SDR method normally obtain a suboptimal solution to the phase shift optimization subproblems, which causes that the objective function's value of the problem $\mathcal{P}1$ may not be non-increasing. Hence, in order to ensure the convergence of our algorithm, we only update the phase shift ${\bf{\Theta}}^{*}$ when it can improve the SSR of the system.
\subsection{Overall Algorithm and Analysis}
The overall algorithm is implemented by iteratively solving the three subproblems, which is summarized in Algorithm 1.
\begin{algorithm}[h]
	\caption{Alternating Optimization Algorithm for $\mathcal{P}1$}
	\begin{algorithmic}[1]
		\State {\textbf{Initialization:}} set initial ${{\bm{\theta }}^{(0)}}$, ${{\bf{w}}^{(0)}}$, ${{\bf{V}}^{(0)}}$, $ {{\bf{w}}_\text{R}^{(0)}}$ and $k = 0$.
		\Repeat
		\State Update $k=k+1$;
		\State Solve problem $\mathcal{P}4$ based on SCA algorithm and then update phase shift ${\bm{\theta}}^{(k)}$;
		\State Solve problem $\mathcal{P}3$ to obtain ${\bf{W}}^{(k)}$ and ${\bf{V}}^{(k)}$ based on SCA algorithm and then get ${\bf{w}}^{(k)}$ by EVD;
		\State Update the receive bemforming vector ${{\bf{w}}_\text{R}^{(k)}}$ by (11);
		\Until The fractional increase of the SSR is below a threshold $\epsilon > 0$ or the maximum
		number of iterations $K$ is reached;
		\State \textbf{Output:} $\bf{w}^{*}$, $\bf{V}^{*}$, ${{\bf{w}}^{*}_\text{R}}$ and $\bm{\theta}^{*}$.
	\end{algorithmic}
\end{algorithm}

Based on AO framework, Algorithm 1 will generate a monotonically increasing (at least nondecreasing) SSR sequence \cite{r2,r11}. Hence, our proposed algorithm is guaranteed to converge because the system SSR have a upper bound due to the limited resources.

The complexity of optimizing the receive beamforming vector is ${{\cal O}}({N_{\text{R}}^{3}})$ due to the matrix inversion in (11). The complexity of solving the transmit beamforming vector and AN covariance matrix optimization subproblem is ${{\cal O}}({2I_{1}N_{\text{T}}^{3.5}})$ due to solving the SDP \cite{r16}, where $I_{1}$ is the number of iterations for SCA algorithm. Similarly, the complexity of optimizing the phase shift is ${{\cal O}}({I_{2}(M+1)^{3.5}})$, where $I_{2}$ is the corresponding number of iterations. Hence, the overall complexity of our proposed algorithm is ${{\cal O}}(L(N_\text{R}^3 + 2{I_1}N_\text{T}^{3.5} + {I_2}{(M + 1)^{3.5}}))$, where $L$ is the number of iterations for AO algorithm.
\section{SIMULATION RESULTS}
In this section, numerical simulations are provided to demonstrate the performance advantages of the proposed algorithm. We assume that the locations of BS, RIS, UL user, DL user and Eve are (0m, 0m), (20m,10m), (60m, 0m), (40m, 5m) and (20m, -5m), respectively. We assume that there are $N_\text{T}=4$ transmit antennas and $N_\text{R}=4$ receive antennas at the BS. The large-scale fading is modelled by $PL\left( d \right) = P{L_0}{\left( {d/{d_0}} \right)^{ - \varpi }}$,  where $P{L_0} =  - 30$dB is the path loss at the reference distance ${d_0} = 1$m, $d$ is the distance, and $\varpi $ is the path-loss exponent. The path-loss exponent of the channels related to the RIS is set to 2.5 and that of other channels are set to 4 \cite{r11}. Similar to \cite{r3,r7}, for small-scale fading, Rayleigh fading is assumed for each link. Besides, each entry of residual self-interference matrix $\bf{H}_\text{BB}$ is identically and independently distributed as $\mathcal{CN}(0,\sigma^{2}_{\text{SI}})$, where $\sigma_{\text{SI}}^{2}$ depends on the capability of the SIC techniques \cite{r12}. Other required parameters are set as follows unless specified otherwise: $M=40$, ${P_\text{U}} = 0.1$W, $\epsilon=10^{-3}$, $K=10$, $\sigma^{2}_{\text{SI}}=-100$dB, and $\sigma _\text{B}^2= \sigma _\text{U}^2 = \sigma _\text{D}^2 =  - 90$dBm.

The proposed FD system with RIS and AN (FD-RIS-AN) is compared to the following benchmark schemes:
\begin{itemize}
	\item HD system with RIS and AN (HD-RIS-AN): In this scheme, the UL and DL transmissions are allocated with half of the time as compared to the FD scheme. In addition, both the AN and RIS are adopted to improve the system SSR performance.
	\item FD system with RIS but without AN (FD-RIS-NoAN): The AN is not applied in the system, while only RIS is used to enchance the FD system SSR performance.
	\item FD system with AN but without RIS (FD-AN-NoRIS): The RIS is not deployed in the system, while only the AN is used for guaranteeing the FD system SSR performance.
\end{itemize}
\begin{figure}
	\centering
	\includegraphics[width=2.4in]{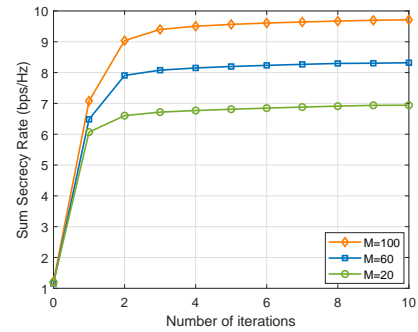}
	\caption{The convergence behavior of proposed algorithm.}
\end{figure}

Fig. 2 investigates the convergence performance of the proposed algorithm with different number of RIS elements. From the figure, it can be seen that as the number of iterations continues to increase, the system SSR increases to convergence in all scenarios, which verifies the effectiveness of the proposed algorithm.
\begin{figure}
	\centering
	\includegraphics[width=2.4in]{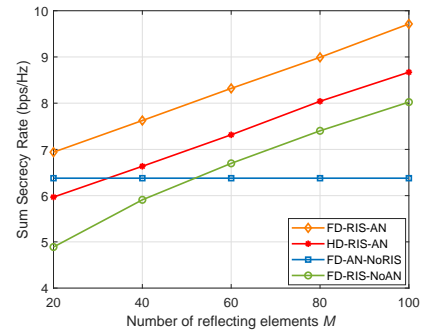}
	\caption{SSR performance versus the number of RIS elements.}
\end{figure}

Fig. 3 illustrates the SSR performance versus the number of reflecting elements. We can find that the system SSR increases with $M$ in the RIS assisted schemes, which indicates that with more reflecting elements, the more degrees of freedom of system design can be exploited and thus achieves higher gains. In addition, the SSR of proposed FD-RIS-AN scheme is always better than that of FD-RIS-NoAN scheme, revealing the effectiveness of AN in improving security performance.

Fig. 4 shows the SSR of different schemes versus the maximum transmit power of the BS. With the increase of transmit power, the SSR of all schemes increase. It is worth noting that the proposed scheme achieves the best performance, which shows the superiority of both AN and RIS assisted FD system in terms of security communications.

The impact of residual SI variance of the BS on the SSR performance is presented in Fig. 5. The SSR performance of all FD schemes decrease with the increase of $\sigma_{\text{SI}}^{2}$, which indicates the powerful SIC methods is necessary for the application of the FD system.
\begin{figure}
	\centering
	\includegraphics[width=2.4in]{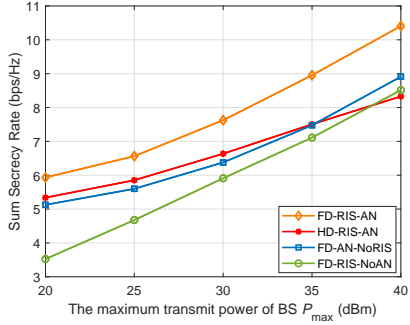}
	\caption{SSR peformance versus the maximum transmit power of BS.}
\end{figure}

\begin{figure}
	\centering
	\includegraphics[width=2.4in]{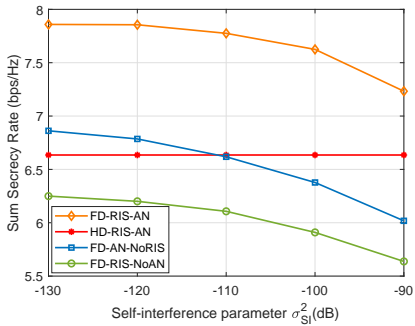}
	\caption{SSR peformance versus the resudial SI variance.}
\end{figure}
\section{CONCLUSION}
This letter studied the SSR maximization for RIS-aided FD systems and the AN is also considered. We propose an efficient algorithm based on AO to optimize the transmit beamforming, receive beamforming, AN covariance matrix at the BS and phase-shifts at the RIS iteratively. Simulation results show the performance advantages of the proposed FD system assisted by both RIS and AN compared with other baseline schemes.
\section*{appendix}
\section*{Proof of Proposition 1}
To prove the Proposition 1, we first rewrite the equivalent problem of $\mathcal{P}3^{\prime}$ as follows
\begin{subequations}\label{opt1}
		\setlength{\abovedisplayskip}{2pt}
	\setlength{\belowdisplayskip}{2pt}
	\begin{alignat}{2} 
	\mathcal{P}:& \mathop {\min }\limits_{{\bf{W}},{\bf{V}},{\eta}_{i}} &\ &{\mu}  \label{opt1A} \\
	& \quad{\textrm {s.t.}}&&- \sum\limits_{i = 1}^4 {{{\log }_2}({\eta _i} + \sigma _i^2)}  + G({\bf{W}},{\bf{V}}|{{\bf{W}}^{(n)}},{{\bf{V}}^{(n)}}) \le \mu,\\
	&&&{\eta _1} \le {P_{\rm{U}}}{\rm{Tr}}({\bf{Q}}{{\bf{H}}_1}) + {\rm{Tr[}}{{\bf{H}}_2}({\bf{W}} + {\bf{V}})],\\
	&&&{\eta _2} \le {P_{\rm{U}}}{\rm{Tr}}({\bf{Q}}{{\bf{H}}_4}) + {\rm{Tr[}}{\bf{h}}_3^H{\bf{Q}}{{\bf{h}}_3}({\bf{W}} + {\bf{V}})],\\
	&&&{\eta _3} \le {P_{\rm{U}}}{\rm{Tr}}({\bf{Q}}{{\bf{H}}_6}) + {\rm{Tr(}}{\bf{h}}_5^H{\bf{Q}}{{\bf{h}}_5}{\bf{V}}),\\
	&&&{\eta _4} \le {\rm{Tr[}}{\bf{h}}_5^H{\bf{Q}}{{\bf{h}}_5}({\bf{W}} + {\bf{V}})],\\
	&&&14(\text{c})-14(\text{e}),
	\end{alignat}
\end{subequations}
where $\sigma _1^2 = \sigma _{\text{B}}^2$, $\sigma _2^2 = \sigma _{\text{D}}^2$ and $\sigma _3^2=\sigma _4^2 = \sigma _{\text{E}}^2$. Problem $\mathcal{P}$ is a convex problem and hence the Slater’s condition holds. Then, the Lagrangian function of $\mathcal{P}$ in terms of $\bf{W}$ can be given by
\begin{equation}
	\setlength{\abovedisplayskip}{2pt}
\setlength{\belowdisplayskip}{2pt}
\begin{array}{l}
	\mathcal{L} = - {\rm{Tr}}({\bf{XW}})+{\lambda _1}{\rm{Tr[}}\nabla _{{{\bf{W}}^{(n)}}}^HG({\bf{W}},{\bf{V}})({\bf{W}} - {{\bf{W}}^{(n)}})] \\ 
	\quad\;\;\;\,- {\lambda _2}{\rm{Tr}}({{\bf{H}}_2}{\bf{W}})
	- {\lambda _3}{\rm{Tr}}({\bf{h}}_3^H{\bf{Q}}{{\bf{h}}_3}{\bf{W}}) - {\lambda _4}{\rm{Tr}}({\bf{h}}_5^H{\bf{Q}}{{\bf{h}}_5}{\bf{W}})\\ 
	\quad\;\;\;\,+ {\lambda _5}\text{Tr}({\bf{W}})+\Delta.
\end{array}
\end{equation}
where ${\lambda _1}$$\sim$${\lambda _5}$ are the corresponding Lagrange multiplier to constraint (21b), (23c), (23d), (23f) and (14c), respectively. Note that $\bf{X}\succeq {\bf{0}}$ is the Lagrange multiplier to positive semi-definite constraint (14d). $\Delta$ in (22) is the collection of terms, which is not relevant for the proof. According to \cite{r14}, we can get the KKT conditions about $\bf{W}$ as folows 
\begin{equation}
	\setlength{\abovedisplayskip}{2pt}
\setlength{\belowdisplayskip}{2pt}
\begin{array}{l}
{{\bf{X}}^*} = {\lambda _5^{*}}{\bf{I}} - {\bf{Y}},
\end{array}
\end{equation}
\begin{equation}
	\setlength{\abovedisplayskip}{2pt}
\setlength{\belowdisplayskip}{2pt}
\begin{array}{l}
{{\bf{X}}^*}{{\bf{W}}^*} = {\bf{0}},
\end{array}
\end{equation}
where ${\bf{Y}} = [{\lambda _2^{*}}{{\bf{H}}_2} + {\lambda _3^{*}}{\bf{h}}_3^H{\bf{Q}}{{\bf{h}}_3} + {\lambda _4^{*}}{\bf{h}}_5^H{\bf{Q}}{{\bf{h}}_5} - \lambda _1^*\nabla _{{{\bf{W}}^{(n)}}}^HG({\bf{W}},{\bf{V}})]$. 

By Sylvester’s rank inequality and (24), we get
\begin{equation}
	\setlength{\abovedisplayskip}{2pt}
\setlength{\belowdisplayskip}{2pt}
\begin{array}{l}
0 = \text{rank}({{\bf{X}}^*}{{\bf{W}}^*}) \ge \text{rank}({{\bf{X}}^*}) + \text{rank}({{\bf{W}}^*}) - {N_\text{T}}.
\end{array}
\end{equation}

Then, we assume $\delta$ denotes the largest eigenvalue of $\bf{Y}$. Based on (23), $\bf{X}^{*}$ is not a positive semi-definite matrix if the $\lambda _5^* \le \delta$, which is contrary to KKT condition. Hence, we obtain $\lambda _5^* \ge \delta$. According to \cite{r14}, the case where multiple eigenvalues have the same value $\delta$ occurs with probability zero, due to the randomness of the channels. Then, we get $\text{rank}({{\bf{X}}^*}) \ge {N_\text{T}} - 1$. According to (25), we have
\begin{equation}
	\setlength{\abovedisplayskip}{2pt}
\setlength{\belowdisplayskip}{2pt}
\begin{array}{l}
\text{rank}({{\bf{W}}^*}) \le {N_\text{T}}{\rm{ - }}\text{rank}({{\bf{X}}^*}){\rm{ = 1}},
\end{array}
\end{equation}
which thus completes the proof.
\ifCLASSOPTIONcaptionsoff
  \newpage
\fi

\bibliographystyle{IEEEtran}
\bibliography{IEEEabrv,myre}

\end{document}